# Enhancing credit risk scorecards with transactional-based behavioural features

Dian Prinsloo[1], Tanja Verster[1,2], Neill Smit[1,2*]

[1]Centre for Business Mathematics and Informatics & Unit for Data Science and Computing, North-West University, South Africa. [2]National Institute for Theoretical and Computational Physics (NITheCS), Potchefstroom, South Africa.

*Corresponding author: neillsmit1@gmail.com / Neill.Smit@nwu.ac.za

## Abstract

**Background:** Credit risk modelling is a fundamental practice in financial institutions, most commonly relying on traditional financial information such as credit history, repayment behaviour, income, and outstanding debt to estimate a borrower's probability of default. Although transactional data are sometimes incorporated through aggregated measures such as recency, frequency, and monetary indicators, behavioural information embedded in raw transaction narratives remains largely unexplored due to its unstructured nature. This represents a missed opportunity to extract richer behavioural insights.

**Aim:** This research addresses a gap in the literature by investigating the use of payment narratives from transactional data as an alternative data source for credit risk modelling.

**Setting:** The study uses transactional data from a large South African financial institution focused on small and medium enterprises.

**Method:** Techniques such as natural language processing and regular expressions are used to transform unstructured transaction descriptions into meaningful, explainable behavioural features by categorising trans-actions into interpretable spending categories. A novel categorisation algorithm was developed and evaluated for classification accuracy. The extracted features are incorporated into a credit risk scorecard to assess their predictive value.

**Results:** The categorisation algorithm demonstrates high classification accuracy. Using these features in the scorecard resulted in a significant increase in Gini compared to a scorecard without them.

**Conclusion:** Text-based transactional information can be systematically extracted, categorised, and operationalised to enhance credit risk model performance.

**Contribution:** The study develops a novel, integrated, explainable methodology for transforming trans-actional text into behavioural features, introduces a structured RegEx-based categorisation approach, and demonstrates their predictive value within traditional scorecard frameworks.

# 1 Introduction

Modelling the probability of default (PD) is a common practice in financial institutions to determine borrower credit risk. Typically, traditional financial data is used in classification models to predict this probability, including credit repayment history, income, outstanding debt, and account balances. Transactional data is sometimes used to identify recency, frequency, and monetary (RFM) spending habits (Birant, 2011). However, transactional spending behaviour across specific categories (e.g. groceries or fuel) is not commonly used, as it depends on large, text-based transaction narratives that are difficult to utilise in raw form. Effective credit risk management is critical for financial institutions, as borrower default can lead to significant financial losses. The PD is one of the most widely used measures of credit risk.

Credit scoring techniques are typically categorised into traditional and non-traditional approaches. Traditional data includes credit history, repayment capacity, capital, and outstanding debt (Ahmed and Rajaleximi, 2019), while non-traditional data includes social, mobile, payment, and location information (Ruiz, 2017). Wang et al. (2018) explore the use of transactional data, showing how bank card transactions can improve credit scoring. Transactional data represents a growing research area with potential to enhance predictive performance. Hlongwane et al. (2024) note that such non-traditional data are often high-dimensional and complex, but can broaden financial inclusion and improve model accuracy, highlighting a gap in the literature.

This paper addresses this gap by exploring payment narratives from transactional data to improve credit risk scorecards, using small and medium enterprises (SME) data from a South African financial institution. The focus is on behavioural credit risk modelling for customers with transactional histories.

Methods for categorising large volumes of raw transactional data are investigated using text mining, natural language processing (NLP), and regular expressions (RegEx). The first objective is to develop an effective categorisation approach, and the second is to evaluate the contribution of categorised transactional features to credit risk scorecards. This study makes three contributions. First, it proposes an explainable methodology to transform raw transactional text into meaningful behavioural features suitable for scorecards. Second, it introduces a structured and explainable categorisation approach using frequency-monetary analysis, domain knowledge, and hierarchical RegEx rules. Third, it empirically evaluates the predictive value of categorised transactional data in improving credit risk scorecards.

The paper is structured as follows. Section 2 reviews literature on traditional and alternative data, text mining, and NLP. Section 3 describes the methodology, including data acquisition, text-mining categorisation, feature engineering, and scorecard evaluation. Section 4 presents the results on categorisation accuracy and scorecard performance. The paper concludes with limitations and future research directions.

# 2 Literature Review

Credit scoring refers to evaluating the creditworthiness of a borrower using statistical models (Abdou and Pointon, 2011). These models, such as logistic regression (LR), are commonly used to predict the PD. Modelling credit risk accurately is important for financial institutions, as it enables effective risk management, fairness, and explainability, since failure in these areas can result in regulatory, legal, and reputational risk (Bahlool et al., 2026). The distinction between traditional and non-traditional (alternative) data is a key concept throughout this study. Table 1 summarises the main differences between these data types. Alternative data varies across borrower types. SME data is typically transaction-driven (e.g. payment narratives), whereas individual and corporate data differ. This study focuses on transactional alternative data, which is widely available for SMEs (Ruiz, 2017).

Traditional credit scoring relies on financial and demographic variables such as credit lines, debt-to-income ratio, repayment history, and account balances (Çalli and Coşkun, 2021; Abdou and Pointon, 2011). Bureau data is also widely used (Hlongwane et al., 2024). These variables reflect past financial behaviour and remain dominant due to availability, explainability, and regulatory compliance (Siddiqi, 2017). The Basel Committee on Banking Supervision (1999) highlight data limitations and model validation as key challenges, which traditional data helps mitigate. Overall, while traditional financial data remains dominant due to its availability, regulatory acceptance, and explainability, its reliance on historical credit behaviour limits its effectiveness for borrowers with incomplete or non-representative credit histories.

Table 1: Traditional versus non-traditional data in credit scoring.

| | **Traditional Data** | **Non-Traditional Data (alternative data)** |
|---|---|---|
| **Definition** | Financial information traditionally used by financial institutions to assess credit risk. | Information not commonly used in standard credit scoring but potentially informative for assessing credit risk. |
| **Non-exhaustive examples** | Payment history, account conduct, in-come information, savings information, lending information, demographic information. | Transactional data, payment narratives, text-data, social media information, mobile data, geographical data, psychometric data. |
| **Data structure** | Structured, well-defined, and numerical data. | Often unstructured or semi-structured, text-based, large, and complex data. |
| **Focus of assessment** | Financial stability, historical borrowing, and repayment patterns. | Behavioural patterns, spending habits, and alternative indicators of creditworthiness. |

Alternative data has gained importance due to digitalisation (World Bank Group, 2019). Sources include utility, psychometric, social, mobile, and transactional data, although not all are applicable to SMEs. Nevertheless, online transactional data is widely applicable for SMEs in financial institutions. Zhang (2025) highlight that alternative data captures behavioural patterns and is often available in real time, enabling customer seg-mentation and tailored products. However, it is complex and often unstructured, requiring advanced processing. Studies show that alternative data improve predictive performance (see, for example, Roeder, 2021). Miliūnaitė (2023) finds significant benefits for SMEs, particularly from payment behaviour data. Alternative data may also im-prove model fairness. Djeundje et al. (2021) note that traditional scoring excludes individuals with limited credit history. Other studies highlight gender bias and data-driven discrimination (Lee and Yang, 2024; Huang, 2024; Hurlin et al., 2024). Malysheva (2024) concludes that alternative data improves performance but involves fairness trade-offs. Consequently, alternative data offer significant potential to enhance prediction accuracy, fairness, and customer inclusion – particularly for SMEs – but its complexity and unstructured nature necessitate advanced, explainable data processing techniques before it can be effectively integrated into credit risk models.

Text mining extracts useful information from large, unstructured data (Shekhar, 2021). NLP transforms ambiguous text into structured data (Gupta et al., 2020; Pajukoski, 2020) using preprocessing steps such as tokenisation and stemming. Shu et al. (2024) show that NLP can enhance credit scoring by incorporating textual data. Advanced methods such as the bidirectional encoder representations from transformer (BERT) model improve text understanding but lack interpretability (Sanz-Guerrero and Arroyo, 2025; Xia et al., 2023). RegEx provides a rule-based and interpretable method for extracting structured information (Yin et al., 2020). This highlights that text mining techniques such as NLP and RegEx are critical enablers for converting unstructured textual data into structured, interpretable features that can be meaningfully incorporated into credit risk scorecards.

Given the literature reviewed, it is evident that alternative data is increasingly used to enhance credit risk modelling beyond traditional financial variables. However, within the broad category of alternative data, existing research has focused mainly on sources such as psychometric data, mobile application data, and social or utility data, with comparatively limited attention given to transactional spending data, particularly text-based payment narratives. Although transactional data are widely available to financial institutions and have been shown to contain rich behavioural information, limited research has explored how unstructured transactional text can be systematically transformed into explainable features for credit risk scorecards, especially in the context of SMEs. Furthermore, while NLP techniques are well-established for processing unstructured text their combined use with rule-based approaches such as RegEx remains under-explored in credit risk modelling. This highlights a clear research gap in understanding how transactional payment narratives can be processed using NLP and RegEx to generate interpretable, behaviour-based features that improve the predictive power of credit risk scorecards. This paper addresses this gap by developing interpretable methods to extract behavioural features from transactional payment narratives to improve credit risk scorecards.

# 3 Methodology

The objective of the methodology is first to categorise raw transactional data into explainable groups, and then to use these categorised results to create behavioural features that can be evaluated in credit risk scorecards. The methodology for this study is illustrated in the flow chart in Figure 1.

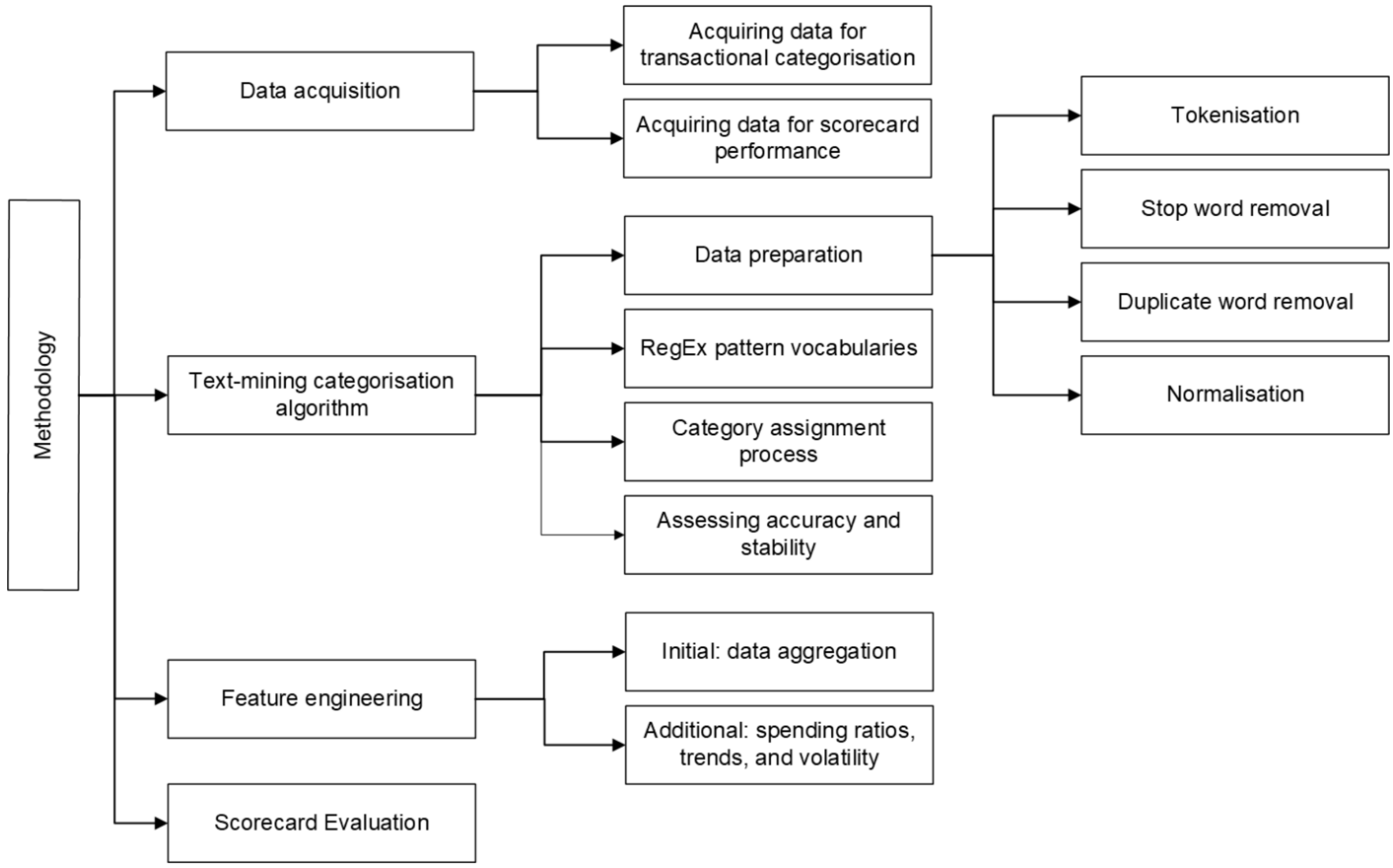

Figure 1: Flow diagram of the methodology structure.

## 3.1. Data acquisition

The data acquisition process consists of two parts. First, raw transactional data was acquired for the categorisation process, and second, data was acquired for the scorecard evaluation.

Data collection for the categorisation process involves acquiring anonymised transactional data from SME clients of a large financial institution in South Africa. This data consists of daily income and expense transactions made by South African businesses, where two fields from this data source is used. The first field is a descriptive text field that provides descriptive information for each transaction in the form of detailed text. This information is used for the text-mining categorisation algorithm. The second field is a numerical field that captures the continuous monetary amount spent or received for each transaction. This data, along with the text data, are used for feature engineering.

For the second objective, evaluating the categorised results on a credit risk scorecard, a good-bad indicator for business clients is used. Credit scores are measured at an observation date, whereas the good-bad indicator is measured at a performance date based on the financial product's performance window. This part of the data is used to establish a baseline performance of the scorecard, against which the additional categorised transactional features can be evaluated.

## 3.2. Text-mining categorisation algorithm

The text-mining categorisation algorithm consists of three main parts: data preparation, the construction of RegEx pattern vocabularies, and the category assignment process. The term *text-mining algorithm* refers to the use of text-mining methods and conditional logic to categorise transactions into groups. In this research, the focus is on text-mining rather than applying formal machine learning (ML) algorithms, such as LLMs. While pre-trained LLMs are trained primarily on general text data in specific languages, such as English, their performance on domain-specific tasks, for instance recognising the names of South African retailers, may be suboptimal.

The transactional data used in this study contain South African-specific retailer names and phrases. For example, transactional data might include phrases related to South African retailers, such as *Checkers* or *Caltex*, which are not commonly used in general English text data. A pre-trained language model might recognise phrases such as *Groceries* or *Fuel* and effectively categorise these phrases into suitable groups, but the majority of the data contains retailer-specific names. Due to this, a geographically tailored categorisation algorithm was developed specifically for the South African retail context.

The raw transactional data is seen as an unsupervised problem, meaning that the data has no predefined labels that can be used for predictions (Buckalew, 2025). Therefore, a simple classification model cannot be trained to predict the categories for the transactions from the raw data. The text-mining algorithm can, however, be used to label the raw data by categorising the transactions, which can then be used as a supervised development base. The text-mining categorisation algorithm was developed using several text-mining packages in Python, specifically `pyspark` (Apache Software Foundation, 2025), `nltk` (NTKL Project, 2026), `re` (Python Software Foundation, 2026), `numpy` (Harris et al., 2020).

The descriptive text field data consist of unclean text entries, which means that they contain unnecessary elements in terms of categorisation into a spending category, such as special characters, numbers, unwanted spaces, and dates. In order to maximise the categorisation efficiency and accuracy, these unnecessary elements are removed as a first data manipulation step to transform the data to a clean and singular form. RegEx-based cleaning techniques are implemented to identify and remove these unwanted characters.

Further data cleaning is performed using a series of NLP-based cleaning techniques. These techniques included tokenisation, stop word removal, duplicate word removal, and normalisation. Tokenisation is the process of converting a sequence of text into smaller parts, known as tokens, which can assist the text analysis process (Awan, 2024). For example, a sequence of text regarding an electronic funds transfer (EFT) transaction such as *"EFT payment to supplier"* is split into the tokens *"EFT", "payment", "to", "supplier".* After tokenisation, words that do not add significant meaning to the text, such as "the", "is", "in", "and", and "to", are discarded during the stop word removal process. Next, normalisation is performed to reduce noise in the data by handling multiple representations exhibited by a single word. Lemmatisation is used for the algorithm in this study, as it retains the actual meaning of the word instead of reducing the word to a root form that does not make sense, as can be the case with stemming (Goyal, 2021). For example, lemmatisation would reduce the word "suppliers" to "supplier", whereas stemming could reduce the word "suppliers" to "supply". Lastly, repeated words are removed during the duplicate word removal process. These NLP-based cleaning techniques are applied recursively until no further changes are made to the text, ensuring that the text is in its most clean and singular form. The final cleaned text is then used as input for the categorisation algorithm.

After the text-cleaning step, a set of vocabularies consisting of RegEx patterns is constructed from the cleaned data. The vocabularies consist of various groups, where a list of RegEx patterns is defined to match specific phrases or words that belong to each group. For example, a *Groceries* category can include RegEx patterns corresponding to South African grocery retailers, such as *CHECKERS, PICK N PAY, SPAR,* and *SHOPRITE*.

In this study, RegEx patterns, vocabularies, and category groups are constructed manually by applying industry-related knowledge of South African retailers and linking this expertise to common phrases found in the transactional data. Less well-known merchant names that are not necessarily present in the data are also assigned to appropriate category groups. Frequency and monetary distribution analyses are performed on each unique phrase to understand its importance and contribution to the data. This process finds phrases that occur more frequently than others and phrases that contribute more to total spending than others, which are then used as a starting point to identify key categories. After constructing the main categories, the vocabularies are expanded using RegEx patterns corresponding to other South African retailers related to these categories, and the remaining infrequent or low-spending phrases are assigned to a category called *Other*.

The RegEx vocabularies can be updated by adding new patterns that emerge from new data or by changing existing patterns to better match phrases to categories, while the process of how the algorithm assigns categories remains constant. After the categories have been created, the algorithm can classify new transactional data based on the text extracted from the descriptive field of the data. The algorithm classifies a phrase into a category by directly assigning the label based on the best-matching RegEx pattern. Consequently, no classification thresholds are required for the classification process. The algorithm first assigns a high-level category, which is then updated by considering additional RegEx pattern matches. For example, a cleaned text observation such as *CHECKERS CHEQUE CARD PURCHASE PAYMENT* may match several patterns. *PAYMENT* or *CARD PURCHASE* may first be assigned to higher-level transaction-type categories, while *CHECKERS* is assigned to a more detailed Groceries category. The algorithm therefore first assigns an interim category and then updates this category until the final, more detailed category is assigned. The category assignment process using this example is also illustrated in Figure 2.

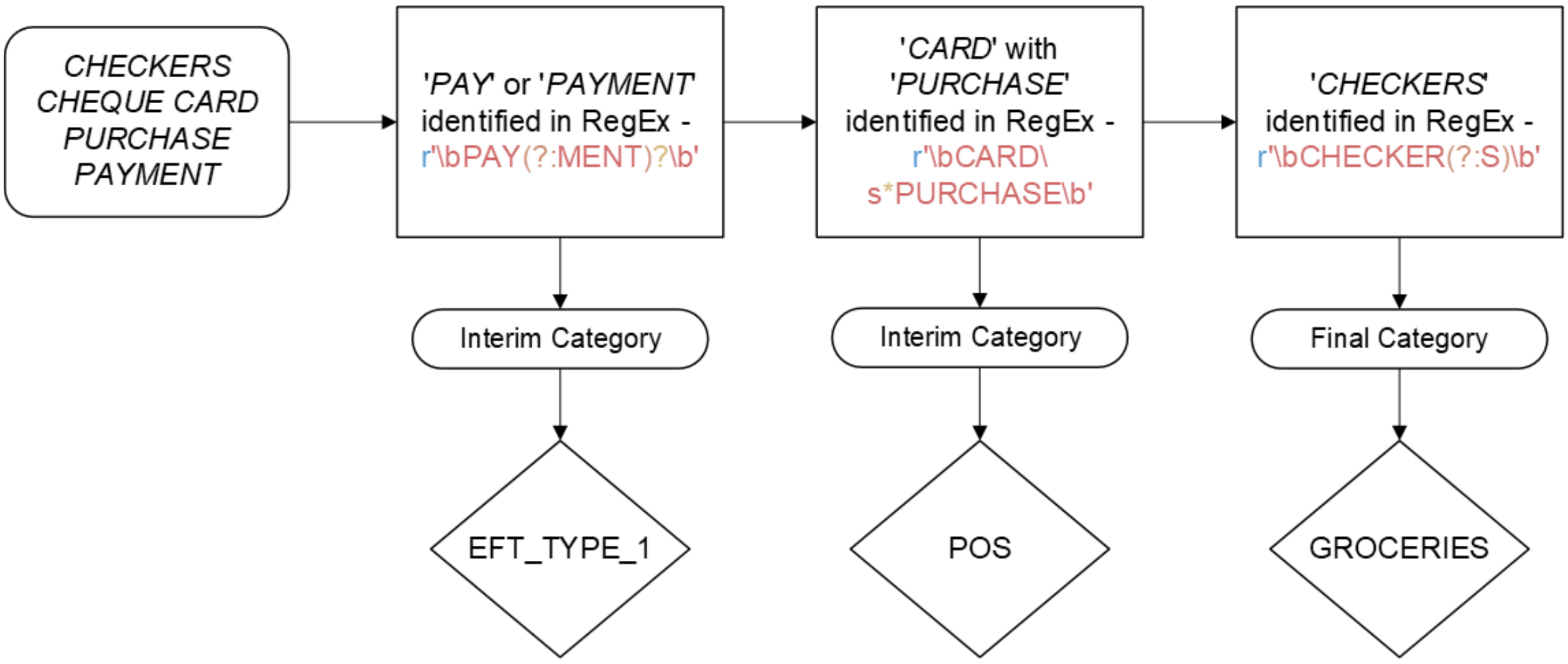

Figure 2: Example of the categorisation process.

### Assessing the accuracy of the algorithm

The performance of the text-mining categorisation algorithm can be evaluated using classification metrics such as precision, recall, and the F1 score. Since there are multiple categories, this creates a multi-class classification problem. For a multi-class problem, the macro-average precision, recall, and F1 score can be computed to assess the accuracy across all categories. Macro averaging assigns equal importance to each class by first calculating the per-class metric and then averaging across all classes (Grandini et al., 2020). An example of a multi-class confusion matrix with $M$ classes is given in Table 2.

As in binary classification, each category has true positives (TP), false negatives (FN), false positives (FP), and true negatives (TN), which are key elements in calculating the classification metrics. The recall for category $k$ measures the ability of the algorithm to correctly identify all transactions that truly belong to category $k$.

Table 2: Confusion matrix for a multi-class problem.

| Categories | | Predicted | | | | |
|---|---|---|---|---|---|---|
| | | $C_1$ | $C_2$ | $C_3$ | ⋯ | $C_M$ |
| Actual | $C_1$ | $C_{1,1}$ | $C_{1,2}$ | $C_{1,3}$ | ⋯ | $C_{1,M}$ |
| | $C_2$ | $C_{2,1}$ | $C_{2,2}$ | $C_{2,3}$ | ⋯ | $C_{2,M}$ |
| | $C_3$ | $C_{3,1}$ | $C_{3,2}$ | $C_{3,3}$ | ⋯ | $C_{3,M}$ |
| | ⋮ | ⋮ | ⋮ | ⋮ | ⋱ | ⋮ |
| | $C_M$ | $C_{M,1}$ | $C_{M,2}$ | $C_{M,3}$ | ⋯ | $C_{M,M}$ |

The precision for category $k$ measures the reliability of the algorithm's predictions when it assigns a transaction to category $k$. The F1 score is the harmonic mean between precision and recall for the category $k$. The recall, precision, and F1 score can be calculated for a category $k$ as

$$Recall_k = \frac{TP_k}{TP_k + FN_k} = \frac{C_{k,k}}{C_{k,k} + \sum_{j=1;\, j\neq k}^{M} C_{k,j}}, \tag{1}$$

$$Precision_k = \frac{TP_k}{TP_k + FP_k} = \frac{C_{k,k}}{C_{k,k} + \sum_{i=1;\, i\neq k}^{M} C_{i,k}}, \tag{2}$$

$$F1\ score_k = \frac{2 \times Precision_k \times Recall_k}{Precision_k + Recall_k}, \tag{3}$$

where:

- $i$ indexes the true category (row index of the confusion matrix).
- $j$ indexes the predicted category (column index of the confusion matrix).
- $\sum_{j=1;\, j\neq k}^{M} C_{k,j}$ represents all samples that truly belong to class $k$ but were misclassified as other classes.
- $\sum_{i=1;\, i\neq k}^{M} C_{i,k}$ represents all samples that were predicted as class $k$ but actually belong to other classes.

After the recall, precision, and F1 score are calculated for each category, the macro-average metrics can be computed and interpreted as overall performance metrics across all categories. As suggested by Grandini et al. (2020), the macro-average metrics can be calculated as

$$Macro\text{-}average\ Recall = \frac{1}{M}\sum_{i=1}^{M} Recall_i, \tag{4}$$

$$Macro\text{-}average\ Precision = \frac{1}{M}\sum_{i=1}^{M} Precision_i, \tag{5}$$

and

$$Macro\text{-}average\ F1\ Score = \frac{1}{M}\sum_{i=1}^{M} F1\ score_i, \tag{6}$$

### Assessing the stability of the algorithm

Model stability is important in financial applications to ensure that the data remains stable over time. This is important in credit risk models to ensure that the predictor variables do not drift over time. The population stability index (PSI) is a common metric used in credit risk to measure the drift in data on a population level (Pathi, 2025). This can be applied on the categorised transactional variable to assess the stability of the multi-class categories. Suppose that a variable is binned into $M$ bins (or categories), then the PSI compares the multinomial frequencies of these bins over two populations or periods. Let $q_i$ denote the relative frequency of the $i$th bin for the base population (say, the development period) and let $p_i$ be the relative frequency of the $i$th bin for the target population (past or future periods). The PSI is then calculated as

$$PSI = \sum_{i=1}^{M} [(p_i - q_i) \times ln(\frac{p_i}{q_i})]. \tag{7}$$

Lewis (1994) proposes the following interpretation for the PSI value:

- $0 \leq PSI < 10\%$ indicates that the target population shows no substantial drift from the base population.
- $10\% \leq PSI < 25\%$ indicates that the target population shows a marginal drift from the base population.
- $PSI \geq 25\%$ indicates that the target population shows a significant drift from the base population.

## 3.3. Feature Engineering

The feature engineering process involves transforming raw data into relevant and explainable information that can be used by models such as LR. When working with large volumes of transactional data, the data need to be converted to the correct format before features can be created. In credit risk modelling, input data typically needs to be on a customer level, meaning that there should be one data entry per customer. Since active customers can have multiple transactions in a given month, the transactional information needs to be aggregated to a customer level per month.

For the purpose of this study, two types of aggregation are considered: monetary and frequency. Monetary aggregation considers the total amount spent by a customer in a given month and frequency aggregation indicates how often a customer made transactions in a given month. These types of aggregation are consistent with common practice in transaction analysis and align with the objective of retrieving specific spending patterns from transactional data.

Additional features are created to give a more sophisticated view of the customer's spending habits. This involves ratios, trends, and volatility measures for a customer's spending behaviour. Ratios provide a relative view of the customer's spending habits, while trends provide a view of how the customer's spending habits change over time. Ratios are calculated by measuring a specific category aggregation relative to the overall spending aggregation, which is given by

$$Ratio = \frac{Aggregated\ Amount\ Spent\ in\ Category}{Total\ Aggregated\ Amount\ Spent} \tag{8}$$

To analyse the spending trend of a customer over time, a linear regression slope is calculated for a customer over a specified period, such as 3, 6, or 9 months. This slope captures the spending behaviour trend at each point in time over the period and is calculated as

$$Slope = \frac{\sum(x_i - \bar{x})(y_i - \bar{y})}{\sum(x_i - \bar{x})^2}, \tag{9}$$

where:

- $x_i$ is the month of the observation, $\forall i = 1,2,\dots,n$.
- $\bar{x}$ is the average of the number of months used.
- $y_i$ is the amount spent in a specific category in month $i$, $\forall i = 1,2,\dots,n$.
- $\bar{y}$ is the average of the amount spent in a specific category over the number of months used.

A volatility measure can also be used to analyse the spending behaviour of a customer over time. The mean standardised volatility can give an indication of how stable or unstable a customer's spending behaviour is over time. It is calculated as the standard deviation of the amount spent in a specific category over a certain period, standardised by the average amount spent in that category over the same period. That is,

$$MSV = \frac{\sqrt{\frac{1}{n-1}\sum(y_i - \bar{y})^2}}{\bar{y}}, \tag{10}$$

where:

- $y_i$ is the amount spent in a specific category in month $i$, $\forall i = 1,2,\dots,n$.
- $\bar{y}$ is the average of the amount spent in a specific category over the number of months used.
- $n$ is the number of months used in the calculation.

Implementing these feature engineering techniques produces a set of features that may describe the spending behaviour of customers while leveraging the categorised data.

## 3.4. Scorecard evaluation

This section discusses the credit risk scorecard used to test the contribution of the newly engineered features from the categorised transactional data. The final step in the methodology is the scorecard evaluation used to assess the results from the text-mining categorisation and feature engineering. At this stage, the data is in a format suitable for a classification model, where text information has been converted into customer-level numerical features. The goal is to evaluate whether the transactional-based behavioural variables improve scorecard performance.

A standard scorecard development process is followed (Siddiqi, 2006). Two scorecards are developed: one excluding the transactional-based behavioural variables (baseline), and one including them. Both scorecards are developed using weight of evidence (WoE) transformations, which map variables into discrete bins to improve interpretability and handling of missing values (Howard et al., 2025). The evaluation involves comparing two models with different sets of independent variables. The first uses only traditional credit risk variables (e.g. turnover, lending, savings, account conduct), while the second combines these with transactional-based behavioural variables. Both models are estimated using LR, which is widely used in credit risk modelling due to its interpretability and explainability. The models predict the PD, which is transformed into a credit score using standard industry scaling approaches (Siddiqi, 2006).

The aim is to evaluate whether the inclusion of transactional-based behavioural variables improves scorecard performance. The primary performance metric used is the Gini coefficient (Laitinen, 2025), derived from the AUC. The baseline scorecard (without behavioural variables) is compared to the extended scorecard, and the difference in Gini coefficients is used to assess performance improvement. The distribution of credit scores (risk concentration) is also assessed using the Herfindahl-Hirschman index (HHI) (Skridulyte and Freitakas, 2012) and is calculated as

$$HHI = \sum_{i=1}^{z} (\%S_i)^2, \tag{11}$$

where:

- $\%S_i$ is the percentage distribution of the $i$th credit score in the population. $\%S_i$ can be computed at the $i$th unique credit score level or over a binned range of credit scores.
- $z$ is the number of unique credit scores in the population.

Lower HHI values indicate a more granular and well-diversified score distribution, reflecting improved risk differentiation. Furthermore, the contribution of variables in each scorecard is evaluated to assess the relative importance of traditional variables and transactional-based behavioural variables.

# 4 Experiment and results

This section presents high-level results for the categorisation algorithm in terms of the accuracy and stability of the algorithm. The predictive performance of credit risk scorecards with and without the newly engineered features is also compared.

## 4.1 Accuracy of the categorisation algorithm

The raw transactional data used to construct and apply the algorithm consist of large volumes of data. To give some scale to the problem, note that the data considered consist of around 40 million monthly transactional observations. This is primarily due to the data including multiple daily transactions per customer per month. To train the categorisation algorithm, a sample of transactional data from 2023 over a 12-month development period is used. The classification performance is then measured on samples consisting of past data (2021, 2022) and more recent data (2024, 2025).

It is not feasible to manually evaluate the accuracy of the algorithm based on all observations. An equal allocation stratified sample approach, considering a total of 34 possible categories, is used to estimate the accuracy of the categorisation algorithm. Samples of size 1000 are drawn, respectively, from the transactions in each year and manually categorised using domain expertise knowledge. The manually assigned (true) categories are then compared to the categories assigned by the algorithm (predicted) to assess the classification accuracy of the algorithm.

The recall, precision, and F1 score are calculated individually for each category $k$ $(k = 1,2,\ldots,M)$ using a one-vs-all approach. There are $M = 34$ categories, where the metric values for each year across the categories are summarised using boxplots in Figure 3. The categorisation algorithm overall achieves a high recall, precision, and F1 score for the majority of categories. It is clear that the algorithm performs slightly worse on the 2021 and 2025 data, but this can be expected since these out-of-time samples are further away from the development period of 2023. Possible reasons for these deviations are that some retailers may have ceased to exist before 2023, meaning they would not be included in the RegEx vocabularies constructed using only 2023 data, or new retailers may have entered the data after 2023, possibly requiring an update to the RegEx vocabularies.

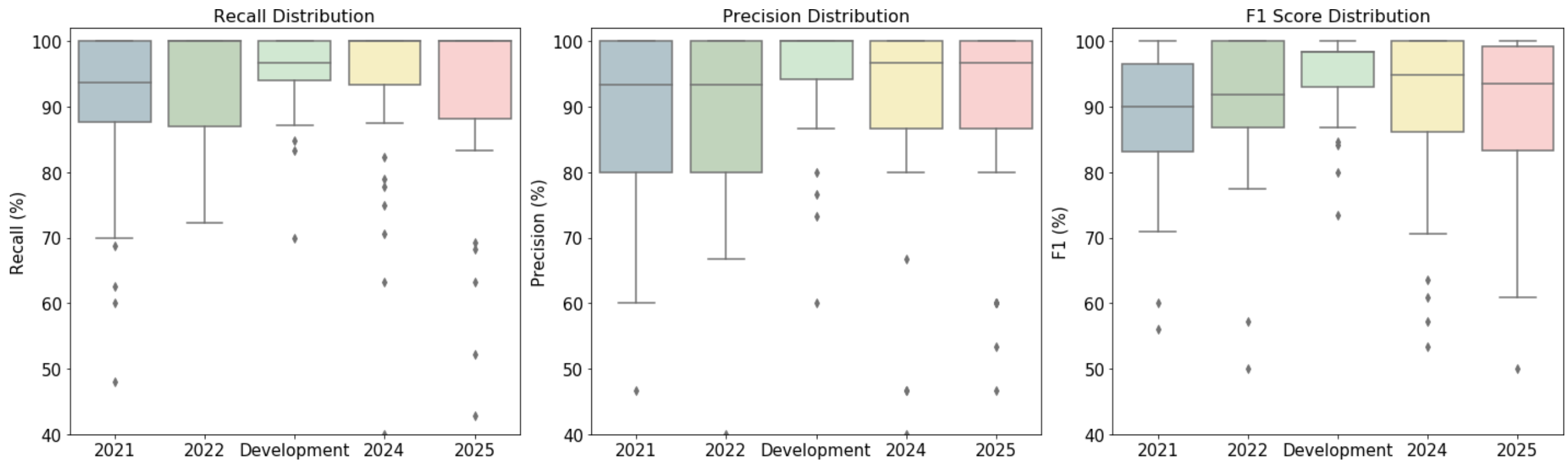

Figure 3: Precision, recall, and F1 score distributions across all categories per year.

Focusing on the more comprehensive classification metric, the full results for the F1 score per category per year are given in Table 3. The classification accuracy in terms of the F1 score is high for most of the categories, while the lowest F1-scores on the out-of-time samples are achieved for the EFT_TYPE_1 category. The EFT_TYPE_1 category is one of the dominating categories, meaning it has many occurrences in the full data, and also the category with the most free-text entries in the data. Free-text entries refer to user-defined text fields in the data used for payment descriptions, which may explain misclassifications for this category. The OTHER category has poor performance for the 2022 and 2024 data. However, as described in the Methodology, this category contains the transactions remaining after assigning the main categories, which may lead to less accurate classification in some periods and better performance on other periods depending on how well the sample data matches the development sample data. The full results for the recall and precision are provided in Table 9 and Table 10.

Using (4), (5), and (6), the macro-average recall, precision, and F1 score are calculated for each year and presented in Table 4. The results highlight the categorisation algorithm's strong overall predictive accuracy, while again indicating slightly less accurate predictions when moving further away from the development period. The metric results per category may be used in future to identify categories in the algorithm which require further refinement, whereas the macro-metric results may be used in future to determine if the algorithm would need to be retrained.

Table 3: F1 score calculated per category per year.

| F1 Score | | | | | |
|---|---|---|---|---|---|
| **Category** | **2021** | **2022** | **2023** | **2024** | **2025** |
| POS | 75.0% | 80.0% | 73.5% | 60.9% | 75.0% |
| EFT_TYPE_1 | 60.0% | 50.0% | 80.0% | 53.3% | 50.0% |
| SERVICE | 75.0% | 78.8% | 84.2% | 70.6% | 63.2% |
| EFT_TYPE_3 | 84.6% | 88.9% | 84.6% | 63.6% | 69.6% |
| TRANSPORT | 88.9% | 88.9% | 86.8% | 92.9% | 60.9% |
| ENTERTAINMENT | 71.0% | 77.4% | 88.5% | 93.3% | 93.8% |
| EQUIPMENT | 82.8% | 89.7% | 88.9% | 85.7% | 70.6% |
| GROCERIES | 85.7% | 87.5% | 90.9% | 84.9% | 90.9% |
| TRANSFERS | 90.3% | 88.9% | 92.9% | 93.3% | 100.0% |
| LOANS_TYPE_1 | 56.0% | 78.6% | 93.8% | 75.0% | 64.3% |
| TAKEAWAYS | 100.0% | 100.0% | 95.1% | 100.0% | 100.0% |
| EFT_TYPE_2 | 88.9% | 88.9% | 96.6% | 92.9% | 92.9% |
| MUNICIPALITY | 93.8% | 90.9% | 96.7% | 96.8% | 100.0% |
| FOREX | 96.6% | 100.0% | 96.7% | 100.0% | 93.3% |
| FUEL | 93.8% | 96.8% | 96.8% | 100.0% | 96.8% |
| SECURITY | 100.0% | 100.0% | 96.8% | 100.0% | 93.8% |
| VEHICLE | 89.7% | 88.9% | 98.3% | 80.0% | 93.3% |
| OTHER | 88.9% | 57.1% | 98.3% | 57.1% | 75.0% |
| SALARY | 96.6% | 100.0% | 98.4% | 100.0% | 96.8% |
| TELECOMS | 96.8% | 100.0% | 98.4% | 96.6% | 92.9% |
| REVERSALS | 88.9% | 100.0% | 98.4% | 96.8% | 93.8% |
| TAX | 96.8% | 100.0% | 98.4% | 100.0% | 96.6% |
| LOANS_TYPE_3 | 76.9% | 85.7% | 98.4% | 88.2% | 81.1% |
| OFFICE | 100.0% | 85.7% | 98.4% | 87.5% | 90.3% |
| LOANS_TYPE_2 | 90.9% | 86.7% | 98.4% | 100.0% | 100.0% |
| INSURANCE | 93.8% | 96.6% | 98.4% | 100.0% | 100.0% |
| PREPAID | 96.6% | 100.0% | 100.0% | 100.0% | 100.0% |
| MEDICAL | 93.3% | 93.3% | 100.0% | 93.3% | 96.6% |
| INTEREST | 80.0% | 92.9% | 100.0% | 92.9% | 92.9% |
| GAMBLING | 93.3% | 100.0% | 100.0% | 100.0% | 96.8% |
| FEES | 96.8% | 100.0% | 100.0% | 100.0% | 100.0% |
| DEBITORDERS | 75.0% | 93.8% | 100.0% | 96.6% | 92.9% |
| PENSION | 100.0% | 100.0% | 100.0% | 100.0% | 100.0% |
| CASH | 87.5% | 96.6% | 100.0% | 100.0% | 100.0% |

Table 4: Algorithm macro-metrics per year.

| Year | Period | Recall | Precision | F1 score |
|---|---|---|---|---|
| 2021 | Past | 90.18% | 87.65% | 87.76% |
| 2022 | Past | 92.36% | 89.80% | 90.07% |
| 2023 | Development | 95.46% | 95.00% | 94.88% |
| 2024 | Recent | 92.44% | 89.80% | 89.77% |
| 2025 | Recent | 90.78% | 88.63% | 88.63% |

## 4.2 Stability of the categorisation algorithm

The classification stability of the text-mining categorisation algorithm is also important and is assessed using the PSI. The PSI is calculated as in (7), where the base population is taken as the development sample, and the past and recent samples are used as the target population. For the purpose of this experiment, the stability of only the categorical variable created by the categorisation algorithm is evaluated. Using the rule of thumb for the PSI as discussed in Section 3.2, one can easily measure how different a target population is compared to the base population. For the purpose of this experiment, the full categorised transactional data is used as a population instead of sampled data.

To monitor the change in PSI, the relative frequencies are calculated for each category on a monthly basis for the target populations. That is, the PSI for each month in 2021, 2022, 2024 and 2025 is calculated using (7), by comparing the data from each month to that of the base population. These results are displayed in Figure 4. The stacked bar graphs illustrate the population distribution of the categorical feature for each month, where the scale is indicated on the left-side vertical axis. The solid black line tracks the monthly PSI, with the PSI scale indicated on the right-side vertical axis. The horizontal yellow and red dotted lines indicate the bands for a marginal and significant shift, respectively, in distribution.

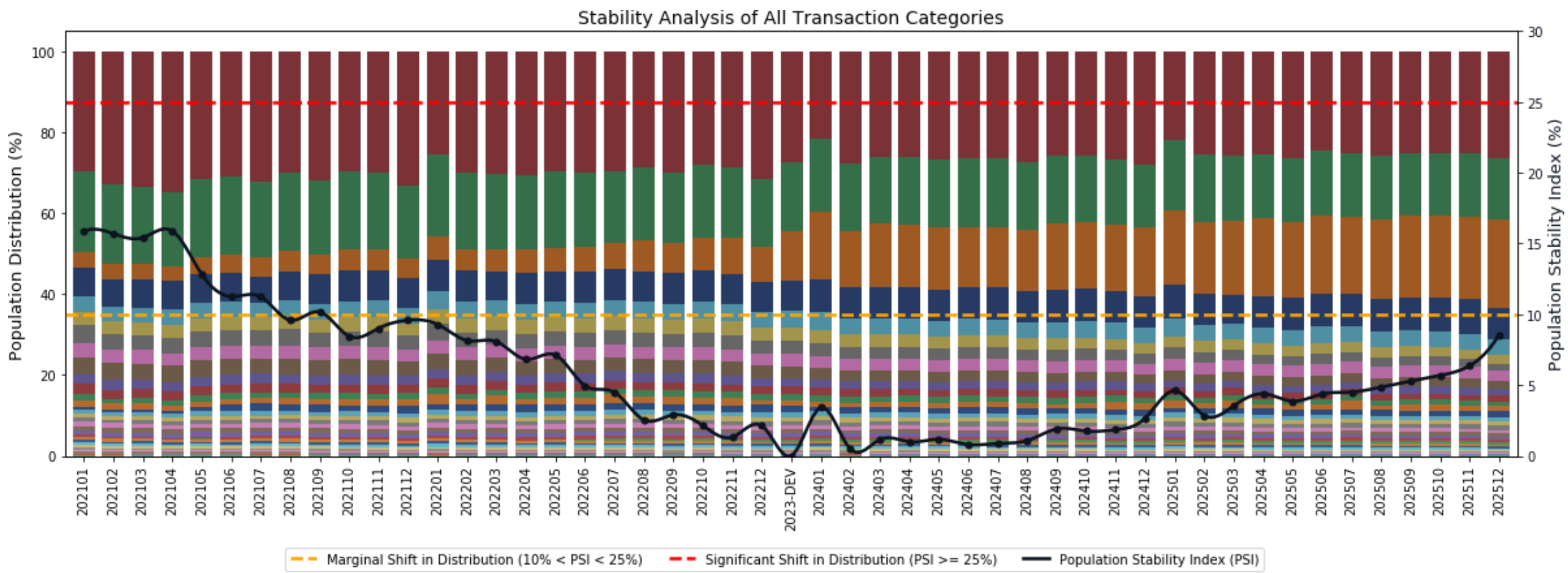

Figure 4: Stability analysis for all transaction categories.

Considering Figure 4, there are no significant changes in the target populations compared to the base population, but there are marginal shifts observed for several months during 2021. The PSI line forms a *U-shape* over the observed out-of-time periods, indicating low PSI values close to development, and higher PSI values for periods further away from development, with more significant changes for the past data than the more recent data. It is understandable that the population drift is more severe on the past data than on the more recent data. This may be because the past data captures transactions made by businesses closer to the COVID-19 world pandemic period. It can be assumed that business expenses looked different during this time than after the pandemic. This is also clear on the left side of Figure 4, where the stability is much more volatile for data at the start of 2021. The stability results are consistent with the accuracy results.

Furthermore, the PSI analysis is repeated where the dominating categories and categories with more free-text entries, such as the EFT, POS, and PREPAID categories, are excluded. The goal of this test is to see if these categories are the main cause of the out-of-time data's shift in PSI. Figure 5 illustrates an updated stability analysis by excluding these dominating categories.

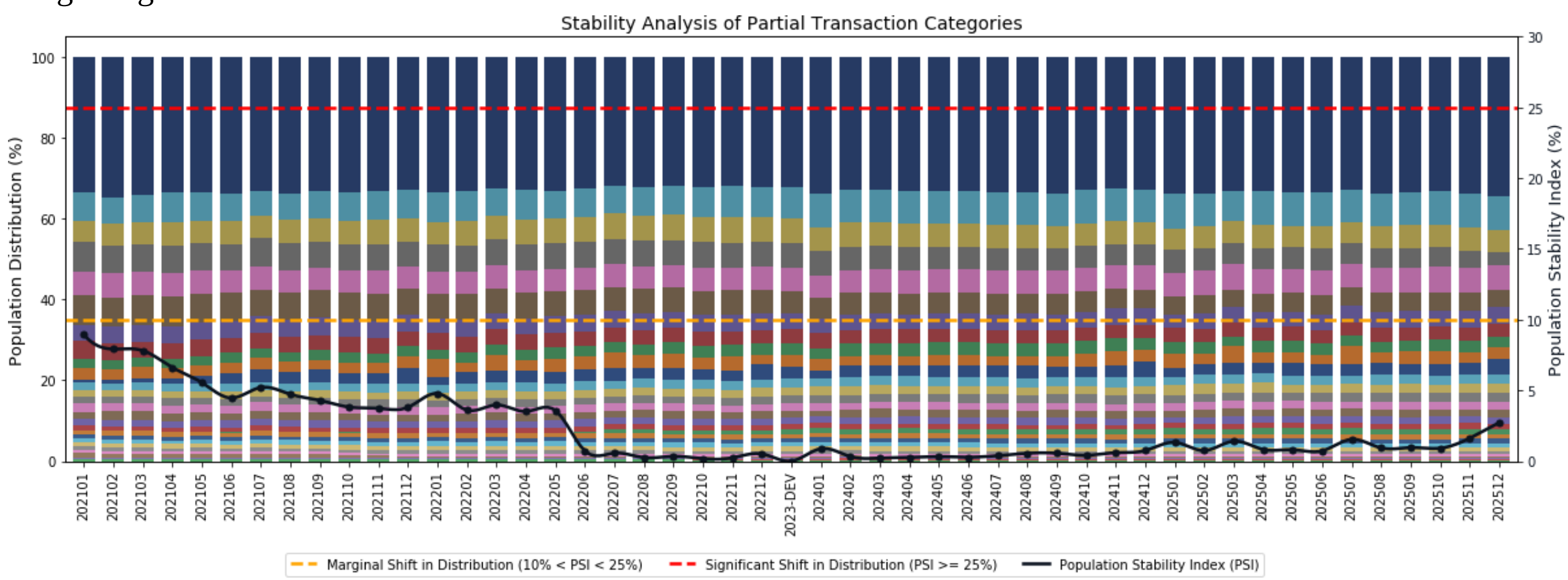

Figure 5: Stability analysis of partial transaction categories for past and present data.

Comparing Figure 4 and Figure 5, it is clear that these dominating categories are the main cause for the PSI shift for the more recent data and has a significant effect for the past data. These are typically the categories where new retailer names or transactions would appear in the data. Table 5 provides the overall PSI when considering the past and recent data, respectively, as a whole. This illustrates a significant drop by excluding the dominating categories.

Table 5: PSI per data group.

| Period | PSI with dominating categories | PSI without dominating categories |
|---|---|---|
| Past data (2021, 2022) | 6.34% | 1.58% |
| Recent data (2024, 2025) | 2.68% | 0.48% |

## 4.3 Comparative scorecard performance

This section focuses on the second main objective of this paper, namely, to test the contribution of the categorised transaction-based behavioural features on a scorecard. Since the scorecard development process is not a focal point of this paper, the methodology discussed in Section 3.4 is brief and the reader is referred to Siddiqi (2006) for further details. The evaluation of the two credit risk scorecards is presented in this section, with emphasis on model performance measured by the Gini coefficient, HHI for risk concentration, and the contributions of transactional-based behavioural features. Two credit risk scorecards are developed using the standard scorecard development process. For a fair comparison, both models used the same development and out-of-time testing periods, the same target definition for identifying a bad customer, and the same number of predictor variables. The only difference is that one model excluded the transactional-based behavioural features (traditional only), while the other included both traditional and transactional-based behavioural features.

Some performance results are arbitrarily adjusted due to confidentiality, but differences reflect true improvements. The model without transactional-based behavioural features is referred to as `Model A` (baseline) and the model including these features as `Model B`. `Model A` produced a baseline Gini coefficient, $Gini_A$, while `Model B` produced $Gini_B$. `Model B` outperformed `Model A`, showing a 14% relative improvement in the Gini coefficient. Table 6 illustrates this using an arbitrary example where $Gini_A = 60\%$.

Table 6: Illustration of Gini coefficient comparison.

| Model | Gini - Illustration |
|---|---|
| `Model A` | 60% ($Gini_A$) |
| `Model B` | 68% ($Gini_B$) |

`Model A` uses 10 traditional features (e.g. turnover, lending, savings, account conduct). `Model B` also uses 10 features, consisting of four traditional features and six transactional-based behavioural features created as discussed in Section 3.3. These features include trends related to EFT_TYPE_1, TAX, CASH, and LOANS_TYPE_1. Table 7 summarises the combined contributions, where contribution percentages are aggregated across the two types of features, showing that transactional-based behavioural features contribute almost 60% of `Model B`.

Table 7: Contributions of feature groups in `Model B`

| Feature Group | Contribution |
|---|---|
| Contribution of traditional variables in `Model B`. | 41.6% |
| Contribution of transactional-based behavioural variables in `Model B`. | 58.4% |

The risk concentration was also compared between the models. A more granular score distribution is preferred. Table 8 shows that `Model B` has a lower HHI, indicating improved granularity.

Table 8: HHI for `Model A` versus `Model B`.

| Model | HHI |
|---|---|
| `Model A` | 1.33% |
| `Model B` | 1.03% |

Although the HHI values are similar, `Model B` shows a more granular distribution, as also illustrated in Figure 6. Figure 6 shows that `Model A` scores are more concentrated around 200, with less spread, while `Model B` scores are more granular and more widely distributed. The red bars represent `Model A`, while the orange and yellow bars represent `Model B`, where orange indicates overlap and yellow indicates non-overlapping regions.

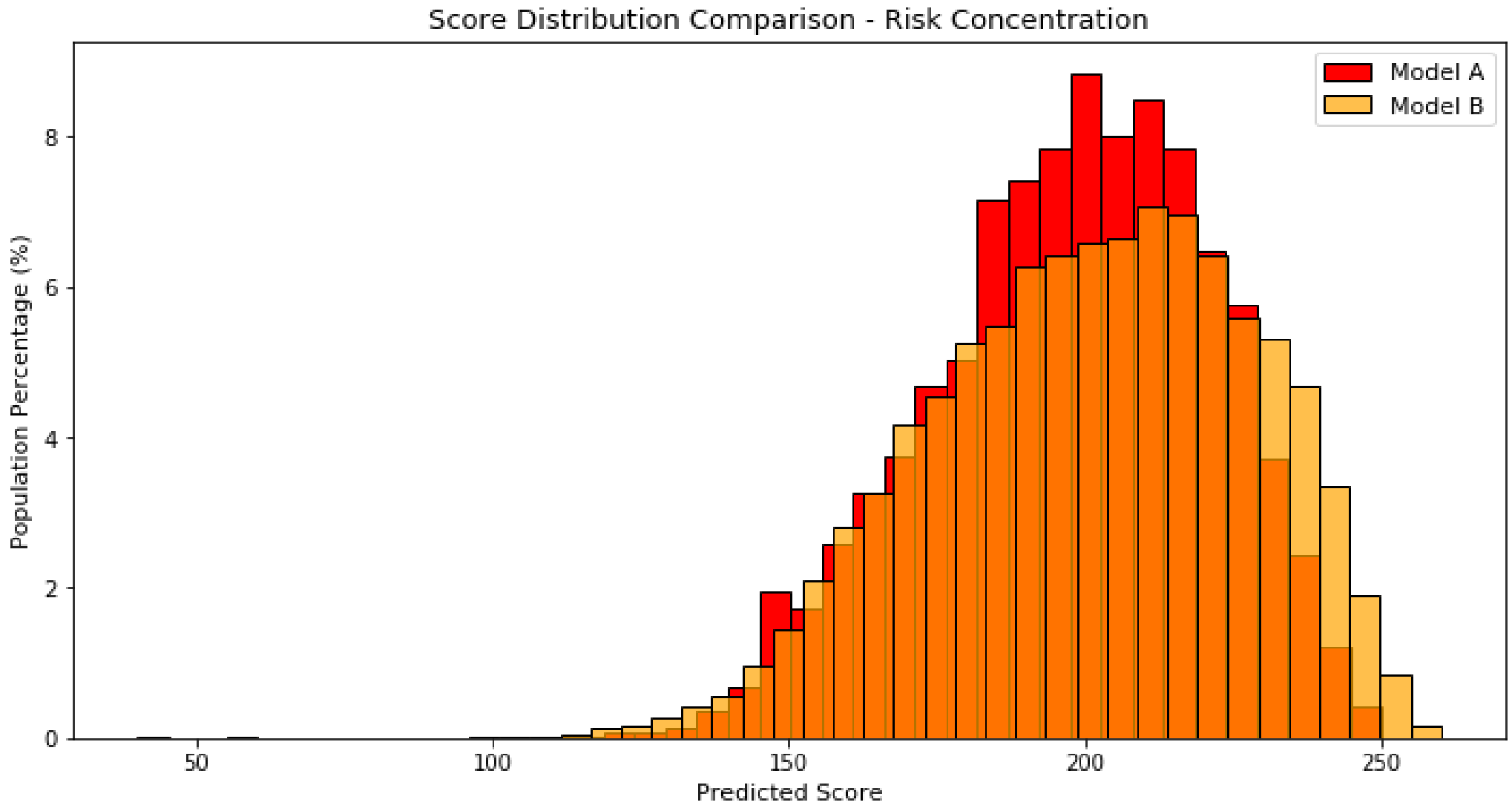

Figure 6: Risk concentration comparison of scorecard with and without transactional-based behavioural features.

The inclusion of categorised transactional-based behavioural variables therefore improves scorecard performance by increasing discriminatory power, as shown by the higher Gini coefficient, and by producing a more granular score distribution. These results suggest that combining financial variables with behavioural spending patterns derived from text-based data can improve the performance of a credit risk scorecard. Transactional behavioural information seems to also capture more dynamic and forward-looking signals not fully reflected in traditional variables, resulting in improved risk differentiation and a more refined ranking of customers.

# 5 Conclusion

This paper investigated how transactional data can be processed and categorised into interpretable spending categories, and whether variables derived from these categories improve the performance of a credit risk scorecard. NLP and RegEx techniques were used to process and categorise transactional text data. While NLP is well established in the text-mining literature (Gupta et al., 2020), the use of RegEx-based approaches for transactional text classification is relatively under-explored. This paper addresses this gap by incorporating and optimising RegEx for the classification of transactional text information.

The results in Sections 4.1 and 4.2 show that the text-mining categorisation algorithm effectively classifies transactions into interpretable categories, achieving a macro-average F1 score close to 90% on a recent sample, indicating balanced performance across classes. The stability of the classification over time is confirmed by a PSI of less than 10%, as shown in Figure 5. Features engineered from these categories capture numerical spending behaviour from text-based information. When evaluated in a credit risk scorecard, the inclusion of transactional behavioural features resulted in a relative improvement of 14% in Gini and contributed approximately 60% of the final model. In addition, these features improved the credit score distribution by reducing risk concentration, as shown in Section 4.3. Overall, the use of alternative data, in the form of a text-mining categorisation algorithm, improved the prediction of credit risk.

These findings contribute to SME credit risk modelling by demonstrating that unstructured transactional text data can be transformed into interpretable behavioural features through an integrated and explainable methodology. The results emphasise the value of a structured and transparent categorisation approach that combines frequency-monetary analysis, expert knowledge, and hierarchical RegEx-based rules, and address a gap in the application of RegEx techniques in transactional text mining. Overall, the results provide empirical evidence that categorised transactional data enhances the predictive performance of traditional credit risk scorecards.

Several limitations are noted. The development of the categorisation algorithm involved human intervention, including reviewing text phrases to identify spending categories, and is therefore unlikely to be fully automated as new patterns emerge. The results show that classification accuracy and stability begin to deviate from development-period benchmarks after approximately two years, suggesting that the algorithm should be peri-odically updated. The algorithm is also specific to a South African transactional and linguistic context, and adaptation to other countries would require additional research. Furthermore, the algorithm was developed using data from customers with existing transactional histories and was evaluated in a behavioural scorecard context, limiting its applicability to origination scorecards without sufficient transactional data.

Future work can extend this research by updating the algorithm using more recent data to improve classification accuracy and stability. In addition, advanced NLP methods such as LLMs and BERT (Kakadiya et al., 2024) could be explored to enhance transactional text processing and classification. Further analysis could also investigate dominant categories such as EFT, POS, and PREPAID, which appear to drive stability patterns observed in Figure 4 compared to Figure 5, with the aim of improving consistent classification over time.

## References

Abdou, H. A. and Pointon, J. (2011). Credit scoring, statistical techniques and evaluation criteria: A review of the literature. *Intelligent Systems in Accounting, Finance and Management*, 18(2-3): 59–88.

Ahmed, M. S. I. and Rajaleximi, P. R. (2019). An empirical study on credit scoring and credit scorecard for financial institutions. *International Journal of Advanced Research in Computer Engineering & Technology*, 8(7): 275–279.

Apache Software Foundation (2025). Pyspark (version 4.1.1) [computer software]. Available at: https://spark.apache.org/docs/latest/api/python/index.html. Accessed: 10 June 2025.

Awan, A. A. (2024). Tokenization in NLP: How it works, challenges, and use cases. DataCamp. Available at: https://www.datacamp.com/blog/what-is-tokenization. Accessed: 16 June 2025.

Bahlool, R., Hewahi, N., and Elmedany, W. (2026). Performance, fairness, and explainability in AI-based credit scoring: A systematic literature review. *Journal of Risk and Financial Management*, 19(2): 104. https://doi.org/10.3390/jrfm19020104

Basel Committee on Banking Supervision (1999). Credit risk modelling: Current practices and applications. Technical Report BCBS 49, Bank for International Settlements, Basel.

Birant, D. (2011). Data mining using RFM analysis. In Funatsu, K. (Ed.), *Knowledge-Oriented Applications in Data Mining*, pp. 91–108. IntechOpen.

Buckalew, A. (2025). Supervised vs unsupervised learning explained. Seldon. Available at: https://www.seldon.io/supervised-vs-unsupervised-learning-explained-2/. Accessed: 08 July 2025.

Çalli, B. A. and Coşkun, E. (2021). A longitudinal systematic review of credit risk assessment and credit default predictors. *SAGE Open*, 11(4). https://doi.org/10.1177/21582440211061333

Djeundje, V. B., Crook, J., Calabrese, R., and Hamid, M. (2021). Enhancing credit scoring with alternative data. *Expert Systems with Applications*, 163: 113766. https://doi.org/10.1016/j.eswa.2020.113766

Goyal, C. (2021). Part 4: Step by step guide to master NLP-text cleaning techniques. Analytics Vidhya. Available at: https://www.analyticsvidhya.com/blog/2021/06/part-4-step-by-step-guide-to-master-natural-language-processing-in-python/. Accessed: 16 June 2025.

Grandini, M., Bagli, E., and Visani, G. (2020). Metrics for multi-class classification: An overview. *arXiv preprint*. https://doi.org/10.48550/arXiv.2008.05756

Gupta, A., Dengre, V., Kheruwala, H. A., and Shah, M. (2020). Comprehensive review of text-mining applications in finance. *Financial Innovation*, 6, 39. https://doi.org/10.1186/s40854-020-00205-1

Harris, C. R., Millman, K. J., Van Der Walt, S. J., Gommers, R., Virtanen, P., Cournapeau, D., et al. (2020). Array programming with NumPy. *Nature*, 585: 357–362.

Hlongwane, R., Ramaboa, K., and Mongwe, W. (2024). Enhancing credit scoring accuracy with a comprehensive evaluation of alternative data. *PLOS ONE*, 19(5): e0303566. https://doi.org/10.1371/journal.pone.0303566

Howard, O., Mwau, C., Kuria Wangui, D., and Mbuya, A. (2025). Credit scoring in digital lending: A stepwise logistic regression and weight of evidence approach in kenya. *The Egyptian Statistical Journal*, 69(2): 424–429.

Huang, Y. (2024). Fairness-aware credit risk assessment using alternative data: An explainable ai approach for bias detection and mitigation. *Artificial Intelligence and Machine Learning Review*, 5(1): 27–29.

Hurlin, C., Pérignon, C., and Saurin, S. (2024). The fairness of credit scoring models. *Management Science*, 72(1): 406–425.

Kakadiya, R., Khan, T., Diwan, A., and Mahadeva, R. (2024). Transformer models for predicting bank loan defaults: A next-generation risk management approach. In *Proceedings of the 2024 IEEE 6th International Conference on Cybernetics, Cognition and Machine Learning Applications (ICCCMLA)*, pp. 26–31, Piscataway, NJ:IEEE.

Laitinen, E. K. (2025). Gini coefficient and AUC in assessing predictive model performance: Effect of ranks. *Theoretical Economics Letters*, 15: 1251–1266.

Lee, J. Y. and Yang, J. (2024). Properties of alternative data for fairer credit risk predictions. *Journal of Data-Centric Machine Learning Research*, 2(2): 1–27.

Lewis, E. M. (1994). *An Introduction to Credit Scoring*. Athena Press, London.

Malysheva, O. A. (2024). The use of alternative data in credit scoring: performance, fairness, and regulatory constraints. *International Journal of Engineering Technology Research & Management*, 8(9): 117–124.

Miliūnaitė, L. (2023). Evaluating the credit risk of SMEs using artificial intelligence, financial and alternative data. Master's thesis, Kaunas University of Technology, Kaunas.

NTKL Project (2026). Nltk (version 3.9.3) [computer software]. Available at: https://www.nltk.org/. Accessed: 10 June 2025.

Pajukoski, J. (2020). Predicting credit rating change using machine learning and natural language processing. Master's thesis, University of Turku, Turku.

Pathi, S. P. (2025). Model evaluation beyond auc: A comparative study of Somers' D, log loss, population stability index (PSI), and kolmogorov–smirnov (KS) statistic in credit risk and healthcare prediction models. *International Journal of Emerging Trends in Computer Science and Information Technology*, 2025: 106–111. https://ijetcsit.org/index.php/ijetcsit/article/view/427

Python Software Foundation (2026). re - regular expression operations. python 3.14 [computer software]. Available at: https://docs.python.org/3/library/re.html. Accessed: 10 June 2025.

Roeder, J. (2021). Alternative data for credit risk management: An analysis of the current state of research. In Pucihar, A. et al. (Ed.), *34th Bled eConference: Digital Support from Crisis to Progressive Change*, pp. 181–194. University of Maribor Press, Maribor, Slovenia.

Ruiz, S. (2017). Credit scoring in microfinance using non-traditional data. Master's thesis, Porto: Universidade do Porto.

Sanz-Guerrero, M. and Arroyo, J. (2025). Credit risk meets large language models: Building a risk indicator from loan descriptions in p2p lending. *Inteligencia Artificial*, 28(75): 220–247.

Shekhar, S. (2021). Text mining and sentiment analysis. *International Research Journal of Engineering and Technology*, 8(1): 282–290.

Shu, M., Liang, J., and Zhu, C. (2024). Automated risk factor extraction from unstructured loan documents: An nlp approach to credit default prediction. *Artificial Intelligence and Machine Learning Review*, 5(2): 10–24.

Siddiqi, N. (2006). *Credit Risk Scorecards: Developing and Implementing Intelligent Credit Scoring*. John Wiley & Sons, Hoboken, NJ.

Siddiqi, N. (2017). *Intelligent Credit Scoring: Building and Implementing Better Credit Risk Scorecards*. John Wiley & Sons, Hoboken, NJ.

Skridulyte, R. and Freitakas, E. (2012). The measurement of concentration risk in loan portfolios. *Economics & Sociology*, 5(1): 51–61.

Wang, H., Zhou, J., and Huang, D. (2018). RFMs method for credit scoring based on bank card transaction data. *Statistica Sinica*, 28(4): 2903–2919

World Bank Group (2019). *Credit Scoring Approaches: Guidelines*. World Bank Group, Washington, DC. Accessed: 30 April 2026.

Xia, Y., Shi, Z., Du, X., and Zheng, Q. (2023). Extracting narrative data via large language models for loan default prediction: when talk is not cheap. *Applied Economics Letters*, 32(4): 481–486.

Yin, C., Jiang, C., Jain, H. K., and Wang, Z. (2020). Evaluating the credit risk of SMEs using legal judgments. *Decision Support Systems*, 136: 113364. https://doi.org/10.1016/j.dss.2020.113364

Zhang, X. (2025). The application of alternative data in credit scoring. In *Proceedings of the International Workshop on Navigating the Digital Business Frontier for Sustainable Financial Innovation (ICDEBA 2024)*, pp. 684–690. Dordrecht: Atlantis Press.

# Appendix: Additional results

Table 9: Recall calculated per category per year.

| Category | 2021 | 2022 | 2023 | 2024 | 2025 |
|---|---|---|---|---|---|
| EFT_TYPE_1 | 48.0% | 37.9% | 70.0% | 40.0% | 42.9% |
| GROCERIES | 92.3% | 82.4% | 83.3% | 77.8% | 83.3% |
| EQUIPMENT | 85.7% | 92.9% | 84.9% | 75.0% | 63.2% |
| ENTERTAINMENT | 68.8% | 75.0% | 87.1% | 93.3% | 88.2% |
| LOANS_TYPE_1 | 70.0% | 84.6% | 88.2% | 70.6% | 69.2% |
| SERVICE | 60.0% | 72.2% | 88.9% | 63.2% | 52.2% |
| TAKEAWAYS | 100.0% | 100.0% | 93.6% | 100.0% | 100.0% |
| SECURITY | 100.0% | 100.0% | 93.8% | 100.0% | 88.2% |
| FUEL | 88.2% | 93.8% | 93.8% | 100.0% | 93.8% |
| POS | 100.0% | 100.0% | 94.7% | 87.5% | 100.0% |
| MUNICIPALITY | 88.2% | 83.3% | 96.7% | 93.8% | 100.0% |
| FOREX | 100.0% | 100.0% | 96.7% | 100.0% | 93.3% |
| TAX | 93.8% | 100.0% | 96.8% | 100.0% | 100.0% |
| INSURANCE | 88.2% | 100.0% | 96.8% | 100.0% | 100.0% |
| TELECOMS | 93.8% | 100.0% | 96.8% | 100.0% | 100.0% |
| LOANS_TYPE_2 | 83.3% | 86.7% | 96.8% | 100.0% | 100.0% |
| SALARY | 100.0% | 100.0% | 96.8% | 100.0% | 93.8% |
| REVERSALS | 100.0% | 100.0% | 96.8% | 93.8% | 88.2% |
| OFFICE | 100.0% | 75.0% | 96.8% | 82.4% | 87.5% |
| LOANS_TYPE_3 | 62.5% | 75.0% | 96.8% | 79.0% | 68.2% |
| PREPAID | 100.0% | 100.0% | 100.0% | 100.0% | 100.0% |
| TRANSFERS | 87.5% | 100.0% | 100.0% | 93.3% | 100.0% |
| CASH | 82.4% | 100.0% | 100.0% | 100.0% | 100.0% |
| OTHER | 100.0% | 100.0% | 100.0% | 100.0% | 100.0% |
| MEDICAL | 93.3% | 93.3% | 100.0% | 93.3% | 100.0% |
| TRANSPORT | 100.0% | 100.0% | 100.0% | 100.0% | 87.5% |
| INTEREST | 100.0% | 100.0% | 100.0% | 100.0% | 100.0% |
| GAMBLING | 93.3% | 100.0% | 100.0% | 100.0% | 93.8% |
| FEES | 93.8% | 100.0% | 100.0% | 100.0% | 100.0% |
| EFT_TYPE_3 | 100.0% | 100.0% | 100.0% | 100.0% | 100.0% |
| EFT_TYPE_2 | 100.0% | 100.0% | 100.0% | 100.0% | 100.0% |
| DEBITORDERS | 100.0% | 88.2% | 100.0% | 100.0% | 100.0% |
| PENSION | 100.0% | 100.0% | 100.0% | 100.0% | 100.0% |
| VEHICLE | 92.9% | 100.0% | 100.0% | 100.0% | 93.3% |

Table 10: Precision calculated per category per year.

| Category | 2021 | 2022 | 2023 | 2024 | 2025 |
|---|---|---|---|---|---|
| POS | 60.0% | 66.7% | 60.0% | 46.7% | 60.0% |
| EFT_TYPE_3 | 73.3% | 80.0% | 73.3% | 46.7% | 53.3% |
| TRANSPORT | 80.0% | 80.0% | 76.7% | 86.7% | 46.7% |
| SERVICE | 100.0% | 86.7% | 80.0% | 80.0% | 80.0% |
| TRANSFERS | 93.3% | 80.0% | 86.7% | 93.3% | 100.0% |
| ENTERTAINMENT | 73.3% | 80.0% | 90.0% | 93.3% | 100.0% |
| EFT_TYPE_1 | 80.0% | 73.3% | 93.3% | 80.0% | 60.0% |
| EFT_TYPE_2 | 80.0% | 80.0% | 93.3% | 86.7% | 86.7% |
| EQUIPMENT | 80.0% | 86.7% | 93.3% | 100.0% | 80.0% |
| VEHICLE | 86.7% | 80.0% | 96.7% | 66.7% | 93.3% |
| MUNICIPALITY | 100.0% | 100.0% | 96.7% | 100.0% | 100.0% |
| TAKEAWAYS | 100.0% | 100.0% | 96.7% | 100.0% | 100.0% |
| FOREX | 93.3% | 100.0% | 96.7% | 100.0% | 93.3% |
| OTHER | 80.0% | 40.0% | 96.7% | 40.0% | 60.0% |
| SECURITY | 100.0% | 100.0% | 100.0% | 100.0% | 100.0% |
| SALARY | 93.3% | 100.0% | 100.0% | 100.0% | 100.0% |
| REVERSALS | 80.0% | 100.0% | 100.0% | 100.0% | 100.0% |
| PREPAID | 93.3% | 100.0% | 100.0% | 100.0% | 100.0% |
| TAX | 100.0% | 100.0% | 100.0% | 100.0% | 93.3% |
| TELECOMS | 100.0% | 100.0% | 100.0% | 93.3% | 86.7% |
| PENSION | 100.0% | 100.0% | 100.0% | 100.0% | 100.0% |
| CASH | 93.3% | 93.3% | 100.0% | 100.0% | 100.0% |
| MEDICAL | 93.3% | 93.3% | 100.0% | 93.3% | 93.3% |
| LOANS_TYPE_2 | 100.0% | 86.7% | 100.0% | 100.0% | 100.0% |
| LOANS_TYPE_1 | 46.7% | 73.3% | 100.0% | 80.0% | 60.0% |
| INTEREST | 66.7% | 86.7% | 100.0% | 86.7% | 86.7% |
| INSURANCE | 100.0% | 93.3% | 100.0% | 100.0% | 100.0% |
| GROCERIES | 80.0% | 93.3% | 100.0% | 93.3% | 100.0% |
| GAMBLING | 93.3% | 100.0% | 100.0% | 100.0% | 100.0% |
| FUEL | 100.0% | 100.0% | 100.0% | 100.0% | 100.0% |
| FEES | 100.0% | 100.0% | 100.0% | 100.0% | 100.0% |
| DEBITORDERS | 60.0% | 100.0% | 100.0% | 93.3% | 86.7% |
| OFFICE | 100.0% | 100.0% | 100.0% | 93.3% | 93.3% |
| LOANS_TYPE_3 | 100.0% | 100.0% | 100.0% | 100.0% | 100.0% |